\documentclass[12pt]{article}
\input colordvi.sty
\usepackage{doi}  
\usepackage[pdftex]{graphicx} 
\usepackage{fancyhdr}
\usepackage{datetime}
\usepackage{cleveref}
\usepackage{amsfonts}
\usepackage{pstricks,pst-node,pst-text,pst-3d}
\usepackage{amsmath,amssymb,amsthm}
\usepackage{url}
\usepackage{hyperref}
\usepackage{amssymb}
\usepackage{color}

\usepackage{cleveref}
\usepackage{amsmath,amsfonts,amssymb,amscd,latexsym,lineno,epsfig}
\input colordvi.sty
\usepackage{graphicx}
\usepackage[simplified]{pgf-umlcd}
\usepackage{etoolbox}

\patchcmd{\endclassAndInterfaceCommon}
{{\umlcdClassInterface}}
{\umlcdClassInterface}
{}{}

\patchcmd{\endclassAndInterfaceCommon}
{{\umlcdClassAbstractClass}}
{\umlcdClassAbstractClass}
{}{}

\newcommand{\N}{{\mathbb N}}

\newcommand{\be}{\begin{equation}}
\newcommand{\en}{\end{equation}}

\begin{document}
\numberwithin{equation}{section}

%
%
%
\begin{center}
\Large{{\bf Abelian versus non-Abelian  B\"acklund Charts: \\ \leftline{some remarks} \hfill 
}}
\end{center}
\normalsize
\begin{center}{
{\bf\large Sandra Carillo},
\\{\rm Dipartimento Scienze di Base e Applicate
    per l'Ingegneria \\  \textsc{Sapienza} - Universit\`a di Roma,  16, Via A. Scarpa, 00161 Rome, Italy} \\ 
\& \\ {\rm National Institute for Nuclear Physics (INFN), Rome, Italy \\ Gr. Roma1, IV - Mathematical Methods in NonLinear Physics}
\\ \vspace{0.2cm}
    {\bf\large Mauro Lo  Schiavo}\\{\rm Dipartimento Scienze di Base e Applicate
    per l'Ingegneria \\    
  \textsc{Sapienza} - Universit\`a di Roma,  16, Via A. Scarpa, 00161 Rome, Italy} \\ \vspace{0.2cm}
     {\bf \large Cornelia Schiebold}\\ {\rm Department of  Science Education and Mathematics, \\
   Mid Sweden University, S-851 70 Sundsvall, Sweden}\\ 
\& \\ {\rm Instytut Matematyki, Uniwersytet Jana Kochanowskiego w Kielcach, Poland}\\  \vspace{0.2cm}
   \today}
 \end{center}

\medskip
\begin{abstract}  
Connections via B\"acklund transformations among different nonlinear evolution equations are investigated aiming to compare corresponding Abelian and non Abelian results. Specifically, links, via B\"acklund transformations, connecting Burgers and KdV-type hierarchies of nonlinear evolution equations are studied. Crucial differences as well as notable similarities between B\"acklund charts in the case of the Burgers - heat equation, on one side and KdV -type equations are considered. 
The B\"acklund charts constructed in \cite{MATCOM2017} and \cite{JMP2018}, respectively, to connect Burgers and KdV-type hierarchies of  operator nonlinear evolution equations show that the structures,  in the non-commutative cases, are richer than the corresponding commutative ones. 
\end{abstract}

\noindent


%

\section{Introduction}
\label{introd}

B\"acklund  transformations in soliton theory are well known to represent a key tool both for finding solutions to applicative problems as well as for investigating symmetry and structural properties admitted by nonlinear evolution equations, see  \cite{Ablowitz, CalogeroDegasperis, RogersShadwick, RogersAmes, RogersSchief, Gu-book}
where  B\"acklund and Darboux Transformations and applications to
 partial differential equations admitting soliton solutions are studied.
The focus of the results in this study is on  B\"acklund transformations as a tool to connect different nonlinear evolution equations both in the Abelian as well as in the non- Abelian setting. Non-commutative  nonlinear equations are considered wherein the unknown is an operator on a~Banach space. The investigation was initiated   by Marchenko \cite{Marchenko} and, later,  further developped in \cite{Aden-Carl, Carl Schiebold}. The obtained results are part of a research project which on   operator equations which takes its origin in \cite{Carillo:Schiebold:JMP2009} wherein, further to other results,  connections, via B\"acklund transformation, among  operator potential Korteweg-de Vries (pKdV),  Korteweg-de Vries (KdV)  and modified  Korteweg-de Vries (mKdV) equations are studied and extended  to the corresponding hierarchies. 
Subsequent developments are comprised in \cite{Carillo:Schiebold:JMP2011} where the special case in which the operator unknowns can be represented as matrices, is considered; solutions are also constructed. Indeed, a great interest on non-Abelian nonlinear evolution equations is testified by a wide bibliography, see  \cite{Olver:Sokolov, LRB, Ku, KK, Hamanaka1, Hamanaka2}. 

Then, 
aiming to construct a an operator counterpart of the B\"acklund  chart in 
 \cite{Fuchssteiner:Carillo:1989a}, the previous B\"acklund chart is further extended in \cite{SIGMA2016, JMP2018} together with the investigation of properties enjoyed by KdV-type 
 non-Abelian equation and the corresponding hierarchies generated on application of the constructed recursion operators. Some of these hierarchies were already known in the literature, while some other ones are new. In particular, in \cite{JMP2018},   new 
 non-commutative hierarchies are constructed on the basis of results in previous works  \cite{SIGMA2016} and by Athorne and Fordy \cite{AF}.  
These hierarchies, {\it mirror} to each other, when commutativity is assumed, reduce to the nonlinear  equation for the KdV eigenfunction, 
({\it KdV eigenfunction} equation)  \cite{boris90} which takes its origin in the early days of {\it soliton} theory when the {\it inverse spectral transform} (IST) method was introduced  \cite{MGK}. It was later studied in 
\cite{boris90, russi, russi2} together with many further nonlinear evolution equations. Notably, in the commutative case, the Dym equation is connected to all the other KdV-type nonlinear evolution equations, via reciprocal transformation \cite{RogersNucci, Fuchssteiner:Carillo:1989a}. Indeed, reciprocal transformations, according to the pioneering work \cite{RogersWong}, are closely connected to the IST method. 
Investigations aiming to classify 
{\it integrable} nonlinear evolution equations are comprised in \cite{Faruk1, Calogero1985, Mikhailov-et-al, Mikhailov-et-al2, Wang}. 

\medskip
{{The material is organised as follows. The opening Section \ref{background} provides the definition of B\"acklund transformation and few properties needed throughout the subsequent parts. 
 Section \ref{burgers} is devoted to connections involving the Burgers equation, both in the Abelian as well as in the  non-Abelian case. The three subsections are concerned, in turn, to the  
 well known commutative Burgers equation   where some of its properties are recalled, to the two non-commutative counterparts of Burgers equation \cite{SIMAI2008, Carillo:Schiebold:JNMP2012, MATCOM2017} and the last subsection points out differences and similarities between the two Abelian and non-Abelian cases. 
 In the next Section \ref{KdV} the attention is focussed on the wide  B\"acklund charts involving  
 KdV-type equations. Also  Section \ref{KdV} is organised in a way  similar to the previous one. Thus, the B\"acklund chart in \cite{BS1, new} is recalled in the subsection devoted to the classical commutative setting. The following subsection is concerned about the B\"acklund chart, constructed in \cite{Carillo:Schiebold:JMP2009} and later extended \cite{SIGMA2016, JMP2018}. The last subsection is concerned about pointing out differences and analogies between the two cases.  }}
\section{Background notions}
\label{background}
Some background notions required in the presented study, are  collected in this section. 
A wide  literature, such as the books \cite{Ablowitz, CalogeroDegasperis, RogersShadwick, RogersAmes, RogersSchief, Gu-book}, to restrict the list to those ones more closely connected with the present investigation, are concerned about B\"acklund transformations and their applications to {\it soliton equations}. 
In this section, for sake of brevity, only definitions,  not uniquely given  in the literature,  are shortly recalled. In particular, the definition of B\"acklund Transformation given by Fuchssteiner \cite{Fuchssteiner1979} and by Fokas and Fuchssteiner  \cite{FokasFuchssteiner:1981},  is adopted.
Consider two non linear evolution equations 
\begin{equation}\label{1}
u_t = K ( u ) ~~~~\text{and} ~~~ v_t = G (v) ~~K, G : M 
\rightarrow TM~,~~ K, G\in C^{\infty}
\end{equation}
where the unknown functions $u, v$ depend both on the independent variables $x,t$ and, for fixed 
$t$, $u (x,t) \in M$, which denotes a {\it smooth} manifold,   then its generic 
{\it fiber} $T_uM$, at $u\in M$, can be identified 
with $M$ itself. In addition, 
when {\it soliton solutions} are considered, $M$ is assumed to coincide, 
for each fixed $t$, with 
the Schwartz space $S$ \footnote{Specifically, the Schwartz space $S$ of {\it rapidly decreasing functions} on ${{\mbox{R\hskip-0.9em{}I \ }}}^n$, is defined as
$S({{\mbox{R\hskip-0.9em{}I \ }}}^n):=\{ f\in C^\infty({{\mbox{R\hskip-0.9em{}I \ }}}^n) : \vert\!\vert f \vert\!
\vert_{\alpha,\beta} < \infty, \forall \alpha,\beta\in \N_0^n\}$, where 
$\vert\!\vert f \vert\!\vert_{\alpha,\beta}:= sup_{x\in{{\mbox{R\hskip-0.9em{}I \ }}}^n} \left\vert x^\alpha D^\beta f(x)
\right\vert $, and  $D^\beta:=\partial^\beta /{\partial x}^\beta$.} of {\it rapidly decreasing functions} on 
${{\mbox{R\hskip-0.9em{}I \ }}}^n$.  In detail, 
\begin{eqnarray}\label{eq.s}
u_t &= K ( u ),~~~ K : M_1 \rightarrow TM_1,~~~{{ u  :(x,t)    \in{{\mathbb R}} 
 \times  {{\mathbb R}}\to u (x,t)   \in M_1}}\\
v_t &= G (v),~~~G : M_2  \rightarrow TM_2,~~~v  :(x,t)    \in{{\mathbb R}} 
 \times{{\mathbb R}}  \to   v (x,t)   \in  M_2 ,
\end{eqnarray}
where $M:= M_1\equiv M_2$, is usually assumed. 
Then,  \cite{FokasFuchssteiner:1981}\,   a B\"acklund transformation can be defined as follows.

\medskip\noindent
{\bf{ Definition}} {\it Given two evolution equations, $ u_t = K (u)$ and $v_t = G (v)$, then $\hbox{B (u , v) = 0}$  
represents a B\"acklund transformation between them 
 whenever  given two solutions of such  equations,   respectively, $u(x,t)$ and $v(x,t)$,  such that 
\begin{equation}
B (u(x,t), v(x,t)) \vert_{ t=0 } = 0 
\end{equation}
then, it follows 
\begin{equation}
B (u(x,t),v(x,t) )\vert_{t=\bar t} = 0,     ~~\forall \bar t >0 ~, ~~~\forall x\in{\mathbb R}.
\end{equation}}

\noindent
The connection between solutions of the two equations via the B\"acklund transformation $B$
 can be graphicallly represented as %
\begin{eqnarray}\label{BC1}
\boxed{u_t = K (u)} \,{\buildrel B \over {\textendash\textendash}}\, \boxed{ v_t = G (v) }~~.
\end{eqnarray}
Among the many properties, note that hereditariness  is preserved under B\"acklund transformations. That is, according to \cite{FokasFuchssteiner:1981, Olver}, let $\Phi (u)$ denote the hereditary recursion operator  admitted by one of the nonlinear evolution equations  (\ref{1}),   say $u_t = K ( u )$,  so that it  can be written as
\begin{equation}\label{base_u}
u_t = \Phi( u ) u_x ~~~\text{where}~~~ K (u) = \Phi (u) u_x~,
\end{equation}
then, also the other equation   (\ref{1}) admits a hereditary recursion operator which
can be constructed via the B\"acklund transformation and the recursion operator 
$ \Phi( u )$. Specifically,  the recursion operator $\Psi( v )$  admitted by the second equation  is given by
\begin{equation}\label{transf-op} 
\Psi(v)=  \Pi\, \Phi (u)\, \Pi^{ -1}  ~~\Longrightarrow~~ G(v) = \Psi (v)\, v_x ~~\text{and}~~ v_t = \Psi( v ) v_x
\end{equation}
where
\begin{equation}\label{pi}
 \Pi : = -B_v^{ -1} B_u~~,~~
          \Pi : T M_1  \rightarrow T M_2~,
\end{equation}
and $B_u$ and $B_v$ denote the Frechet derivatives of the B\"acklund transformation $B(u,v)$. Then, according to \cite{FokasFuchssteiner:1981}, 
on subsequent applications of the admitted recursion operators, 
 the result can be extendeed to the two generated  hierarchies
 \cite{Fuchssteiner1979} 
\begin{equation}
\displaystyle u_{t} =\left[\Phi (u)\right]^{n} u_x ~~~~~\text{and} ~~~~~~ 
v_{t }= \left[\Psi (v)\right]^{n} v_x~,~~ n\in\N ~~;
\end{equation}
 their {\it base members} 
equations, which correspond to $n=1$, coincide with  equations (\ref{1}). Fixed any $m\in\N$, the two equations $u_{t} =\left[\Phi (u)\right]^{m} u_x$ and $v_{t }= \left[\Psi (v)\right]^{m}$ are connected  via 
the same B\"acklund Transformation  which connects the two base members equations.
This extension  is graphically represented, $\forall~n\in\N$, as 
\begin{equation}\label{BT-hier}
\boxed{u_{} = \left[\Phi (u)\right]^{n} u_x }{\,\buildrel B \over {\textendash\textendash}}\,
\boxed{ v_{t} \ =\  \left[\Psi (v)\right]^{n} v_x} ~.   
\end{equation}

 \section{Burgers B\"acklund charts}\label{burgers}
 
This section is devoted to the Burgers equation. It represents the most simple example of nonlinear evolution equation whose linearisation via a B\"acklund transformation, the well known Cole-Hopf transformation \cite{Cole:1951, Hopf:1950}, represented a crucial step. Indeed, the Cole-Hopf transformation opened the way to the study of many applicative problems, such as initial boundary value problems, see for instance \cite{Guo:Carillo, CalogeroDeLillo}.
 This section is divided in three different subsections which, respectively, are concerned about the Abelian, the non-Abelian Burgers equation and, finally, a comparison between some properties in the two cases.
 
 Specifically, the first subsection, provides a brief survey of some properties enjoyed by the Burgers equation and the corresponding hierarchy, in the Abelian framework; accordingly, the unknown is assumed to be a real valued function \footnote{sometimes, for short, the term {\it scalar} is used to distinguish the case of a commutative unknown function with respect to the non-commutative one, which is represented by an operator valued unknown.}.
 
  The next subsection is concerned about the non-Abelian  Burgers equations and the corresponding hierarchies \cite{SIMAI2008, Carillo:Schiebold:JNMP2012, MATCOM2017}. The first appearence of a non-commutative  Burgers  hierarchies is due to Levi,  Ragnisco  and Bruschi \cite{LRB} who were concerned about matrix equations. The  non-commutative Burgers equations and their properties were studied by Kupershmidt in \cite{Ku}, the  hereditariness of the recursion operator of the non-commutative Burgers equation  in  \cite{SIMAI2008} was proved in   \cite{Carillo:Schiebold:JNMP2012}. The   same recursion operator was  independently obtained, via a Lax pair method in  \cite{GKS}, by G\"urses, Karasu and Turhan \cite{GKT}.  Recently, in \cite{MATCOM2017}, where a comparison with the mentioned results is provided,  also the 
 {\it mirror} non-Abelian Burgers equation, together with the admitted  recursion operator and, hence, the corresponding hierarchy are constructed. 
 
 In the last subsection there is a concise comparison between the two different cases Abelian versus non-Abelian.
  \subsection{Abelian B\"acklund chart}\label{burgers1}
  Burgers equation reads:
  \begin{equation}\label{burgers}
v_t =  v_{xx }+ 2  v v_{x} 
\end{equation}
It is linked via  the Cole-Hopf transformation \cite{Cole:1951, Hopf:1950}, which can be represented as the special B\"acklund transformation
\begin{equation}\label{C-H}
CH:~~~~~~~~~B(u,v)=0~, ~~~ \text{where}~~~ B(u,v)= v u-u_x~,
\end{equation}
to the  linear heat equation
\begin{equation}\label{heat}
u_t =  u_{xx }~. 
\end{equation}
The latter admits the (trivial) hereditary recursion operator $\Phi (u)=D$, where 
$D$ denotes derivation with respect to the spatial variable $x$. Hence,  the heat equation reads
  \begin{equation}\label{base_u}
u_t = \Phi( u ) u_x ~~~\text{where}~~~ K (u) = \Phi (u) u_x~, ~  \Phi (u)=D
\end{equation}
and the corresponding hierarchy, respectively, can be written as
 \begin{equation}\label{hier_u}
u_t =[ \Phi( u )]^n u_x ~~~\text{where}~~~ \Phi (u)=D~, ~  n\in\N~.
\end{equation}
As well known \cite{Fuchssteiner1979}  the link \eqref{C-H} between the heat \eqref{heat} and the Burgers  \eqref{burgers}  
equations can be depicted via the following B\"acklund chart 
\begin{eqnarray}\label{BC1}
\boxed{u_t =  u_{xx }} \,{\buildrel {CH} \over {\textendash\textendash}}\, \boxed{ v_t = v_{xx } +
2 v v_{x} }~~. 
\end{eqnarray}
The link via Cole-Hopf transformation
 \cite{Cole:1951, Hopf:1950}, on one side, allows to construct solutions of assigned initial boundary value problems  modelled via Burgers equation,  as  testified by many results among which  also \cite{Guo:Carillo, CalogeroDeLillo}. On the other side, indicates how to obtain, on application of formula \eqref{transf-op}, from   the recursion operator of the heat equation, the Burgers recursion operator 
\begin{equation}\label{Burgers-op}
\Psi(v)=  D + v_x D^{ -1} +v ~.
\end{equation}
Furthermore, the B\"acklund chart 
can extended to the corresponding hierarchies according to \eqref{BT-hier}, wherein $\Phi(u)=D$ and $\Psi(v)$ 
is given by \eqref{Burgers-op}, that is
\begin{eqnarray}\label{BC2}
\boxed{u_t = [ \Phi( u )]^n u_x} \,{\buildrel {CH} \over {\textendash\textendash}}\, \boxed{ v_t = [\Psi(v)]^n v_{x} }~~, ~  n\in\N~.
\end{eqnarray}
As a consequence, according to the wide literature, see for instance \cite{CalogeroDegasperis, Fuchssteiner1979}, also the infinitely many admitted conserved quantities can be constructed. 
 \subsection{non-Abelian B\"acklund chart}\label{burgers2}
 This subsection is concerned about the non-Abelian Burgers equations which both represent the non-Abelian counterpart of the Abelian Burgers equation. 
 For notational convenience, 
from here on, operator unknown are denoted via capital letters while lower case letters are used referring to real valued ones so that, for instance, the unknown functions  $u$, in the linear heat and $v$ in the Burgers  equations in Subsection \ref{burgers1} are denoted by lower case letters, while, in the present subsection capital case letters are used since operator equations are considered.

 Consider the linear heat equation
 \begin{equation}\label{Heat}
U_t=K(U) ~~, ~~ K(U) = U_{xx}
\end{equation}
where, now, the unknown is an operator according to the general theory in \cite{Aden-Carl, Carl Schiebold, Carillo:Schiebold:JNMP2012}. 
 The two different Cole-Hopf transformations  \cite{MATCOM2017}, namely the
\begin{equation}\label{B}
B_1(U,S)=0~, ~~~ \text{where}~~~ B_1(U,S)= US-U_x~~~ \Longrightarrow~~ S=U^{-1}U_x,
\end{equation}
and its {\it mirror}
 \begin{equation}\label{B2}
B_2(U,R)=0~, ~~~ \text{where}~~~ B_2(U,R)= RU-U_x~~~ \Longrightarrow~~ R=U_x U^{-1},
\end{equation}
give, respectively, the non-Abelian  Burgers equation
\begin{equation}\label{bu}
S_t=G_1(S) ~~, ~~ G_1(S) = S_{xx}+2S S_x,
\end{equation}
and the non-Abelian  Burgers mirror equation
\begin{equation}\label{mbu}
R_t=G_2(R) ~~, ~~ G_2(R)  = R_{xx} + 2R_x R~.
\end{equation}
The recursion operator admitted by the heat equation is represented by the trivial operator $\widehat\Phi(U)=D$, the non-Abelian Burgers and the mirror  non-Abelian Burgers equations \cite{Carillo:Schiebold:JNMP2012, MATCOM2017} are, respectively,  the operators 
$ \Psi(S)$ and $\Phi(R)$ which, when $L_{W}$ and $R_{W}$ denote, in turn,  left and right multiplication by ${W}$ for any $W$, are given as follows
\begin{equation} \label{rec RE}
      \Psi(S) = (D+C_S) (D + L_S) (D+C_S)^{-1} ,~~~ \text{where}~~~C_S:=\big[S, \cdot\big],
\end{equation}
which can also be written as \cite{Carillo:Schiebold:JNMP2012}
\begin{equation} \label{rec}
   \Psi(S)=D + L_S + R_{S_x} (D+C_S)^{-1};
\end{equation}
and
\begin{equation}\label{mrec}
\Phi(R)= (D-C_R)(D+R_R)(D-C_R)^{-1}, ~~~ \text{where}~~~C_R:=\big[R, \cdot\big],
\end{equation}
The corresponding hierarchies, namely, in turn, the heat, the non-Abelian Burgers and the 
{\it mirror} non-Abelian Burgers  hierarchies are represented as
\begin{equation} \label{hierarchies}
   U_{t_n} = D^{n-1}U_x, ~~~  S_{t_n} = \Psi(S)^{n-1}S_x, ~~~  R_{t_n} = \Phi(R)^{n-1}R_x, \quad n\geq 1.
\end{equation}
The lowest order  members of the first two hierarchies are, respectively
\begin{equation}\label{heat-buh}
\begin{array}{ccc}
   U_{t_1} &= U_x, ~~~~~~~~~~~~~~~~~~~~~~   & S_{t_1} = S_x, \hfill \\
   U_{t_2} &= U_{xx} , ~~~~~~~~~~~~~~~~~~ & S_{t_2} = S_{xx} + 2SS_x, \hfill \\
   U_{t_3} &= U_{xxx} , ~~~~~~~~~~~~~~~~~~~~~~~ & S_{t_3} = S_{xxx} + 3SS_{xx} + 3S_x^2 + 3 S^2S_x.
\end{array}
\end{equation}
while, the first members of the mirror non-Abelian Burgers hierarchy, are
\begin{equation}\label{mbuh}
\begin{array}{ccc}
  R_{t_1} &=&R_x, \hfill \\
  R_{t_2} &=&R_{xx} + 2r_xR, \hfill \\
  R_{t_3} &=&R_{xxx} + 3r_{xx}R+ 3r_x^2 + 3R_xR^2.
\end{array}
\end{equation}
Notably, in contrast with the non-Abelian Burgers hierarchy,  each member of its mirror
hierarchy can be obtained from the corresponding one by substitution of right
multiplication  by $S$ with left multiplication by $R$. 

The links among them can be summarised in the following B\"acklund chart which, again, can be extended to the whole corresponding hierarchies.
 \begin{figure}[h]

\unitlength1cm
\begin{picture}(8,6.75)
   
   \put(4,5.5){\framebox(2.5,1){\shortstack{\footnotesize linear heat \\ $U_t=U_{xx}$}}}
   
   \put(1.3,3){\framebox(3.5,1){\shortstack{\footnotesize mirror  Burgers\\ \footnotesize $R_t = R_{xx} +2R_xR$}}}
    \put(6,3){\framebox(3.5,1){\shortstack{\footnotesize Burgers \\ \footnotesize  $S_t = S_{xx} +2SS_x $}}} 
 




  
   \put(3.5,4.25){\vector(1,1){1}}		 \put(3.4,4.75){\footnotesize $\hskip-4em  U= R_xR^{-1}$}
   \put(6.5,4.25){\vector(-1,1){1}} 	 \put(6.2,4.75){\footnotesize $ U=  \ S^{-1} S_x$}
   

\end{picture} \vskip-3cm
\caption{Burgers and mirror Burgers equations and their B\"acklund links: the non-com\-mu\-ta\-ti\-ve case.}
\label{fig nc Burgers chart}

\end{figure}
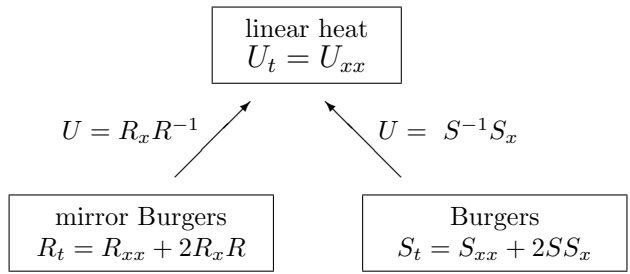

%

 \subsection{non-Abelian  versus Abelian  B\"acklund charts}\label{burgers3}
This brief subsection is concerned about differences and similarities between the Abelian and non-Abelian Burgers hierarchies. 
 Indeed, there are two different non-Abelian transformations which generalise  the Abelian Cole-Hopf transformation. Hence, when the linear heat equation is considered, depending on whether the unknown is a real valued function, say {\it scalar} for simplicity, the usual (Abelian) Burgers equation \eqref{burgers} is obtained. 
 Conversely, if the unknown  in  the linear heat equation is an operator, 
 then the two different non-Abelian equations \eqref{bu} and \eqref{mbu} are obtained. 
 On the other hand, the Cole-Hopf transformation which links the Burgers to the heat equation, both in the non-Abelian  as well as in the Abelian case, allows to construct the recursion operator \eqref{Burgers-op} admitted by the (commutative) Burgers equation, or the two different recursion operators, in turn 
 \eqref{rec RE} and \eqref{mrec}, admitted by the (non-commutative) Burgers and by its mirror  equation, in the non-commutative case. Furthermore, the form of the latter ones is more complicated, however, as expected, when commutativity is assumed, both the Burgers and mirror Burgers recursion operators reduce to the usual (commutative) Burgers recursion operator. These operator allow to generate two hierarchies of nonlinear evolution equations: they are both the same ones obtained by Kupershmidt \cite{Ku} who construct them via a recursive definition of the hierarchies themselves. Also G\"urses, Karasu and Turhan \cite{GKT} obtained the same
recursion operator $\Psi(S)$ and the  hierarchy it generates via a method \cite{GKS} based on the Lax pair formulation. 
The direct proof the hereditary property enjoyed by the recursion operators,  in \cite{MATCOM2017}, requires more involved computation and {\it ad hoc} routines when proved via computer assisted computations to take into account the non-commutativity of products.

The following Section \ref{KdV} shows that    non-Abelian  KdV-type equations are characterised by an
even richer structure.
 \section{KdV-type B\"acklund charts}\label{KdV}
 This Section is devoted to the two different B\"acklund charts connecting KdV-type equations, in turn, in the commutative case and in the non-commutative one. Specifically, to start with, the B\"acklund chart constructed in \cite{BS1, new} is given. Then, the operator B\"acklund chart obtained in \cite{Carillo:Schiebold:JMP2009} and its
   extensions in \cite{SIGMA2016, JMP2018}, are considered. 
   Finally, a comparison between the two different situations closes this section. 
 \subsection{Abelian B\"acklund chart}\label{KdV1}
  In this subsection, the links among the various scalar nonlinear evolution equations are depicted in the B\"acklund chart in \cite{new} wherein, inspired by the non-commutative results in \cite{JMP2018}, 
  the   previous B\"acklund chart constructed in
 \cite{Fuchssteiner:Carillo:1989a} is extended to include also the KdV eigenfunction equation.  
The links,  up to this stage,  are summarised in the following  B\"acklund chart. 

\begin{figure}[h]
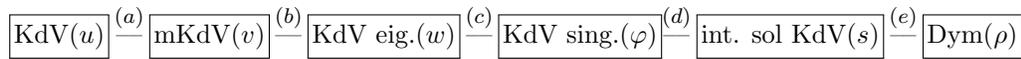

\begin{gather*}
\mbox{$\footnotesize
\boxed{\!\text{KdV}(u)\!}\, {\buildrel (a)
\over{\text{\textendash\textendash}}} \, \boxed{\!\text{mKdV}(v)\! } \, {\buildrel (b) \over{\text{\textendash\textendash}}} \, \boxed{\!\text{KdV eig.}(w)\! } \, {\buildrel (c) \over{\text{\textendash\textendash}}} \, \boxed{\!\text{KdV~sing.}(\varphi)\!}
 {\buildrel (d) \over{\text{\textendash\textendash}}}\, \boxed{\!\text{int. sol KdV}(s) \!} \,
{\buildrel (e) \over{\text{\textendash\textendash}}} \, \boxed{\!\text{Dym}(\rho)\!}\,$}\label{BC1*}
\end{gather*}

\caption{KdV-type B\"acklund chart: the Abelian case.}
\label{fig-BS1-ext}
\end{figure}

\noindent
The third order nonlinear evolution equations in the B\"acklund chart are, respectively 
\begin{alignat*}{3} 
& u_t = u_{xxx} + 6 uu_x \qquad && \text{(KdV)}, & \\
& v_t = v_{xxx} - 6 v^2 v_x \qquad && \text{(mKdV)},& \\
& w_t = w_{xxx} - 3 {{w_x w_{xx}}\over w}\qquad && \text{(KdV eig.)},& \\
& \varphi_t = \varphi_x \{ \varphi ; x\} , \quad \text{where} \ \ \{ \varphi ; x \} :=
 \left( { \varphi_{xx} \over \varphi_x} \right)_x -
{1 \over 2 }\left({ \varphi_{xx} \over \varphi_x} \right)^2 \qquad && \text{(KdV~sing.)}, & \\
& s^2 s_t = s^2 s_{xxx} - 3 s s_x s_{xx}+ {3 \over 2 }{s_x}^3 \qquad && \text{(int.\ sol~KdV)}, & \\
& \rho_t = \rho^{3} \rho_{\xi \xi \xi} \qquad && \text{(Dym)}.&
\end{alignat*}
The B\"acklund transformations  linking these equations among them,  following the order in the B\"acklund chart itself, are:
 \begin{gather}\label{links}
(a) \ \ u + v_x + v^2 =0 , \qquad \qquad \qquad (b) \ \ v -{{ w_{x} \over w} } = 0,\\
(c) \  w^2 -{  \varphi_x}  = 0,\!\qquad \qquad  \qquad\qquad (d) \ \ s - \varphi_x =0, 
\end{gather} 
and 
 the {\it reciprocal} transformation:
\begin{equation}
(e) ~\ {\bar x} : = D^{-1} s (x), ~\rho(\bar x) :=  s(x), ~~~~~\text{where}~~~ ~D^{-1}:= \int_{-\infty}^x d\xi .\label{rec}
\end{equation}
 The latter exchanges the role played by the dependent and independent variables between them; hence,  $\bar x= \bar x(s,x)$ and  $ \rho(\bar x) :=  \rho(\bar x(s,x))$. Details on the reciprocal transformation $(e)$ are analysed in \cite{BS1, Fuchssteiner:Carillo:1989a} while a  
general introduction on reciprocal transformations together with various applications are illustrated in \cite{RogersShadwick}.
 
Notably,  the  B\"acklund chart   represents a useful tool to reveal  the nonlinear evolution equations it links  enjoy invariances, which might be new or already known. 
In particular, the invariance $M$ under the M\"obius group of transformations exhibited by the KdV singularity manifold equation (KdV \-sing.), allows to further extend the B\"acklund chart as  indicated in the following Fig.\ref{fig KdV-ext-chart}.
 \begin{figure}[h]
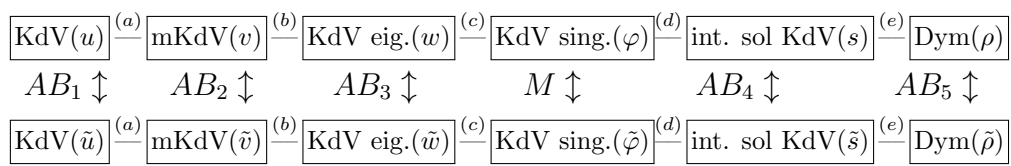


\begin{eqnarray*}
\mbox{\footnotesize $ \boxed{\!\text{KdV}(u)\!}\, {\buildrel (a)
\over{\text{\textendash\textendash}}} \, \boxed{\!\text{mKdV}(v)\! } \, {\buildrel (b) \over{\text{\textendash\textendash}}} \, \boxed{\!\text{KdV eig.}(w)\! } \, {\buildrel (c) \over{\text{\textendash\textendash}}} \, \boxed{\!\text{KdV~sing.}(\varphi)\!}
 {\buildrel (d) \over{\text{\textendash\textendash}}}\, \boxed{\!\text{int. sol KdV}(s) \!} \,
{\buildrel (e) \over{\text{\textendash\textendash}}} \, \boxed{\!\text{Dym}(\rho)\!}\,$}
 \\ \footnotesize 
AB_1\updownarrow~ ~~~ ~~AB_2 \updownarrow~~~~~~~~AB_3 \updownarrow~~~~~~~~~~~M \updownarrow~~~~~~~~~~~~AB_4 \updownarrow~~~~~~~~~~~~AB_5 \updownarrow~~~ \\
\mbox{\footnotesize $\boxed{\!\text{KdV}(\tilde u)\!}\, {\buildrel (a)
\over{\text{\textendash\textendash}}} \, \boxed{\!\text{mKdV}(\tilde v)\! } \, {\buildrel (b) \over{\text{\textendash\textendash}}} \, \boxed{\!\text{KdV eig.}(\tilde w)\! } \, {\buildrel (c) \over{\text{\textendash\textendash}}} \, \boxed{\!\text{KdV~sing.}(\tilde \varphi)\!}
 {\buildrel (d) \over{\text{\textendash\textendash}}}\, \boxed{\!\text{int. sol KdV}(\tilde s) \!} \,
{\buildrel (e) \over{\text{\textendash\textendash}}} \, \boxed{\!\text{Dym}(\tilde \rho)\!}\,$}
\end{eqnarray*}
\caption{Abelian KdV-type hierarchies  B\"acklund chart: induced invariances.}
\label{fig KdV-ext-chart}

\end{figure}
Indeed, the M\"obius invariance allows to obtain the auto-B\"acklund transformations $AB_k, k= 1\dots 5$.  
Thus,  AB$_1$ and AB$_2$ are the well known auto-B\"acklund transformations admitted by the KdV and the mKdV hierarchies \cite{Miura, CalogeroDegasperis, Fuchssteiner:Carillo:1989a}. The auto-B\"acklund transformation   AB$_3\equiv$ denotes 
an invariance enjoyed by the KdV eingenfunction equation \cite{new}.   
Finally,    AB$_4$, and  AB$_5$, respectively, are auto-B\"acklund transformations, enjoyed by, respectively,    the int. sol. KdV and Dym equations, according to \cite{Fuchssteiner:Carillo:1989a}. Notably, the constructed B\"acklund chart not only can be can be extended in $2+1$- dimensions \cite{walsan1} but also can be regarded as a 
{\it constrained} version of such an extension  \cite{walsan2}. Hence, both in $1+1$ \cite{BS4, Guo:Rogers} as well as in $2+1$ dimensions \cite{Rogers:1987}, solutions to  Dym equations problems can be obtained. Furthermore, according to \cite{Fuchssteiner:Carillo:1989a}, the Hamiltonian and bi-Hamiltonian structure of all the nonlinear evolution equations in the B\"acklund chart can be constructed from those ones admitted by the KdV equation 
\cite{Magri,[12], Fuchssteiner:Carillo:1990a, Benno-Walter}. As discussed in \cite{ActaAM2012}, a B\"acklund chart connecting the Caudrey-Dodd-Gibbon-Sawata-Kotera and Kaup-Kupershmidt hierarchies  \cite{CDG, SK, Kawa}  whose base member is a 5th order  nonlinear evolution equations   similar to the one connecting KdV-type equations can be constructed \cite{Rogers:Carillo:1987b, BS1}.
%

All the  links in the B\"acklund chart, in Fig. \ref{fig KdV-ext-chart}, are valid for the whole hierarchies of nonlinear evolution equations generated via application of the admitted  recursion operators. Hence, each element in the Fig. \ref{fig KdV-ext-chart} can be interpreted as the whole corresponding hierarchy. The recursion operator,   respectively, are
{\begin{alignat*}{3} 
 &   \Phi_{\text{KdV}} (u) =D^2+2DuD^{ -1}+2u &&    \text{(KdV), }    \\
 &   \Phi_{\text{mKdV}} (v) =D^2-4DvD^{ -1}vD &&    \text{(mKdV),}      \\
 &    \Phi_{\text{KdV  eig.}} (w) =
{1 \over {2 w}}D w^2 \!\left[ D^2+2 U + D^{ -1} 2 UD\right]\! {1 \over {w^2 }} D^{ -1} \!2w &&    \text{(KdV eig.), }    \\
 &    \Phi_{\text{KdVsing}} (\varphi) = 
\varphi_x\left[ D^2 +  \{ \varphi ; x\}+D^{-1}  \{ \varphi ; x\}D\right] {1\over \varphi_x} ~~~~~~~~~~~~ &&  \text{(KdV sing.), }    \\
 &    \Phi_{\text{KdVsol}} (s) =D s\left[ D^2 +   S + D^{ -1}  SD\right] {1\over s} D^{-1} &&    
 \text{(int.\ sol KdV),}  
\\
 &    \Phi_{Dym} (\rho) =\rho^3D^3 \rho D^{-1}\rho^{-2} &&    \text{(Dym), }  \end{alignat*}}
where 
\begin{equation}
U:=  {w_{xx} \over {w}}- 2 {w_{x}^2 \over {w^2}}~~~~,~~~~S:=  ( {s_x \over s } )_x -{1 \over 2}   ( {s_x \over s}) ^ 2~~.
\end{equation}
and the links among such  hierarchies of nonlinear evolution equations, are indicated in \eqref{links}-\eqref{rec}.

 \subsection{non-Abelian B\"acklund chart}\label{KdV2}
 This subsection is concerned about the B\"acklund chart connecting operator KdV-type equations. Accordingly, the results comprised in \cite{Carillo:Schiebold:JMP2009, SIGMA2016, JMP2018}, are summarised
 in the following picture.
 
 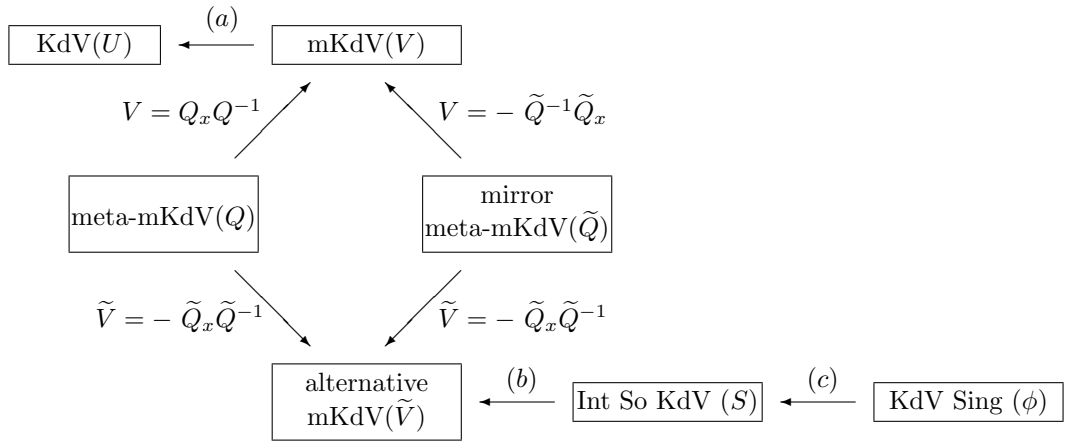
\begin{figure}[h]

\unitlength1cm
\begin{picture}(15,6.75)
   
   \put(0.5,5.5){\framebox(2,0.5){\shortstack{\footnotesize KdV$(U)$}}}
   \put(4,5.5){\framebox(2.5,0.5){\shortstack{\footnotesize mKdV$(V)$}}}
   
   \put(1.3,3){\framebox(2.5,1){\shortstack{\footnotesize \\ \footnotesize meta-mKdV$(Q)$}}}
    \put(6,3){\framebox(2.5,1){\shortstack{\footnotesize mirror \\ \footnotesize meta-mKdV$(\widetilde Q)$}}} 
 
   \put(4,0.5){\framebox(2.5,1){\shortstack{\footnotesize alternative \\ \footnotesize mKdV$(\widetilde V)$}}}
   \put(8,0.75){\framebox(2.5,0.5){\shortstack{\footnotesize Int So KdV $(S)$}}}
   \put(12,0.75){\framebox(2.5,0.5){\shortstack{\footnotesize KdV Sing $(\phi)$}}}




   \put(3.75,5.75){\vector(-1,0){1}}   \put(3.1,6){\footnotesize $(a)$}
  
   \put(3.5,4.25){\vector(1,1){1}}		 \put(3.4,4.75){\footnotesize $\hskip-4em  V= Q_xQ^{-1}$}
   \put(3.5,2.75){\vector(1,-1){1}}	 \put(3.4,2){\footnotesize $\hskip-5em \widetilde V= - \  \widetilde Q_x\widetilde Q^{-1}$}
   \put(6.5,4.25){\vector(-1,1){1}} 	 \put(6.2,4.75){\footnotesize $  V= - \ \widetilde Q^{-1}\widetilde Q_x$}
   \put(6.5,2.75){\vector(-1,-1){1}}    	 \put(6.2,2){\footnotesize $\widetilde V= - \  \widetilde Q_x\widetilde Q^{-1}$}
   
   \put(7.75,1){\vector(-1,0){1}}        \put(7.1,1.2){\footnotesize $(b)$}
   \put(11.75,1){\vector(-1,0){1}}      \put(11.1,1.2){\footnotesize $(c)$}

\end{picture}
\caption{KdV-type  hierarchies  B\"acklund chart: the non-Abelian case.}
\label{fig nc KdVchart}

\end{figure}

\noindent In Fig. \ref{fig nc KdVchart}, the third order nonlinear operator evolution equations \footnote{All the unknown are denoted via capital case letters with the only exception of the KdV sing. equation since $\Phi$ is used to indicate recursion operators.} are, in turn,
\begin{alignat*}{3} 
& U_t = U_{xxx} + 3 \{ U, U_x \} \qquad && \text{(KdV)}, & \\
&  V_t = V_{xxx} - 3 \{ V^2, V_x \} \qquad && \text{(mKdV)},& \\
&   Q_t = Q_{xxx} -3Q_{xx}Q^{-1}Q_{x} && \text{(meta-mKdV)},& \\
&   \widetilde Q_t = \widetilde Q_{xxx} -3\widetilde Q_{x} \widetilde Q^{-1}\widetilde Q_{xx} && \text{(mirror meta-mKdV)},& \\
&   \widetilde V_t = \widetilde V_{xxx} + 3 [ \widetilde V, \widetilde V_{xx} ] -6 \widetilde V \widetilde V_x \widetilde V\qquad && \text{(amKdV)},& \\
&  \phi_t = \phi_{x} \{ \phi; x \}  , \quad \text{where} \ \    \{ \phi ; x \} = \big( \phi_x^{-1} \phi_{xx} \big)_x - \frac{1}{2} \big( \phi_x^{-1} \phi_{xx} \big)^2 \qquad && \text{(KdV~sing.)}, & \\
& S_t = S_{xxx} - \frac{3}{2} \big( S_xS^{-1}S_x \big)_x  \qquad && \text{(int.\ sol~KdV)}. &
\end{alignat*}
while the B\"acklund transformations linking the KdV with the mKdV, the amKdV  with the  int.sol~KdV and the latter  with the   KdV~sing. are,
respectively
\begin{alignat*}{3} 
&   U = -\  (V^2+V_x) \hfill &&  {(a)} & \\
&    \widetilde V = \frac{1}{2} S^{-1} S_x \qquad\qquad\qquad\qquad\qquad\qquad\qquad &&  {(b)}& \\
&      S = \phi_x ~.&&  {(c)}
\end{alignat*}
Again, all the nonlinear evolution equations in the B\"acklund chart admit a recursion operator; hence, the B\"acklund chart can be extended to the whole corresponding hierarchies. Indeed, the recursion operator themselves can be constructed on application of formula \eqref{transf-op} according to \cite{FokasFuchssteiner:1981}. In this, way, in \cite{SIGMA2016, JMP2018} the recursion operators of the newly inserted nonlinear evolution equations in the B\"acklund chart are obtained. The recursion operators, in turn are the following ones.
{\footnotesize{\begin{alignat*}{3} 
 & 
  \Phi_{\rm KdV} (U) = D^2 + 2 A_U + A_{U_x} D^{-1} + C_U D^{-1} C_U D^{-1} &&    \text{(KdV), }    \\ 
 &  \Phi_{\text{mKdV}} (V)= (D - C_VD^{-1}C_V) (D - A_VD^{-1}A_V) &&    \text{(mKdV),}      \\ 
 &  \Phi_{\text{amKdV}} (\widetilde V) = (D+2C_{\widetilde V})(D-2R_{\widetilde V}) (D+C_{\widetilde V})^{-1} (D+2L_{\widetilde V})D (D+C_{\widetilde V})^{-1}  &&    \text{(amKdV), }    \\ 
 &  \Phi_{\text{mmKdV}} (Q) = R_{Q} D^{-1} (D+C_{Q_x Q^{-1}}) (D-A_{Q_x Q^{-1}}D^{-1}A_{Q_x Q^{-1}}) (D-C_{Q_x Q^{-1}})R_{Q^{-1}} ~~~~~ &&  \text{(meta-mKdV), }    \\ 
  &   \Phi_{\text{mmmKdV}} (\widetilde Q) = L_{\widetilde Q} D^{-1} (D-C_{\widetilde Q^{-1}\widetilde Q_x})
	                                  			(D-A_{\widetilde Q^{-1}\widetilde Q_x}D^{-1}A_{\widetilde Q^{-1}\widetilde Q_x})
										   (D+C_{\widetilde Q^{-1}\widetilde Q_x})L_{\widetilde Q^{-1}}  \!\!\!  &&  \text{(mirror meta-mKdV), }    \\ 
&   \Phi_{\text{KdVsing}}(\phi)  
         = L_{\phi_x}  \mathbb{D}^{-1} (\mathbb{D}-A_{N(\phi_x)})
                                                             (\mathbb{D}-C_{N(\phi_x)}\mathbb{D}^{-1} C_{N(\phi_x)})
                                                             (\mathbb{D}+A_{N(\phi_x)}) L_{\phi_x^{-1}}&&    
 \text{(KdV sing.).}  
 \end{alignat*}}
where 
\begin{equation}
\mathbb{D}:=D+C_{\phi}, ~~~C_W:=\big[W, \cdot\big], ~~~A_W:=\big\{W, \cdot\big\} ~~\text{any~arbitrary~} W~,
\end{equation}}
All the recursion operators are hereditary \cite{Carillo:Schiebold:JMP2009, SIGMA2016, 
JMP2018}, therefore, all the links in Fig.  \ref{fig nc KdVchart} hold true for the whole 
corresponding hierarchies. Furthermore,  known invariances allow to prove new ones. 
This is the case of the non-Abelian M\"obius invariance enjoyed by the non-Abelian 
KdV sing. equation and by the corresponding hierarchy which allow to prove nontrivial 
invariances admitted by hierarchies of nonlinear evolution equations in the same 
B\"acklund chart \cite{JMP2018}. In particular, invariances enjoyed by the meta-mKdV and
mirror meta-mKdV hierarchies can be proved \cite{JMP2018}.
 \subsection{non-Abelian  versus Abelian B\"acklund charts}\label{KdV3}
This subsection collects few remarks on the comparison between the Abelian and non-Abelian  
B\"acklund charts connecting KdV- type equations. They, respectively, are depicted in Fig.  
\ref{fig-BS1-ext} and in   Fig. \ref{fig nc KdVchart}. Notably, in the non-commutative case two 
different equations, namely  meta-mKdV and mirror meta-mKdV, 
\begin{alignat*}{3} 
&   Q_t = Q_{xxx} -3Q_{xx}Q^{-1}Q_{x} \hskip4cm&& \text{(meta-mKdV)},& \\
&   \widetilde Q_t = \widetilde Q_{xxx} -3\widetilde Q_{x} \widetilde Q^{-1}\widetilde Q_{xx} && \text{(mirror meta-mKdV)},& 
\end{alignat*}
represent both the non-commutative counterpart of the nonlinear  equation for the KdV eigenfunction, termed, for short, {\it KdV eigenfunction} equation 
$$w_t = w_{xxx} - 3 {{w_x w_{xx}}\over w}~.\hskip6.5cm  (KdV eig.)$$
 The latter, inserted in the commutative B\"acklund chart in \cite{new}, finds its meaning in connection with the Lax pair formulation of the KdV equation \cite{boris90, russi, MGK, russi2}.  Conversely, the non-commutative equations, denoted as meta-mKdV and mirror meta-mKdV, are introduced in the non-commutative B\"acklund chart in \cite{JMP2018}. 
 They both represent novel nonlinear evolution equations. Both these equations  admit  ahereditary recursion operator, constructed in \cite{JMP2018} together with soliton solutions. Notably, the hereditariness of the admitted recursion operator follows from the hereditariness of the KdV recursion operator \cite{Schiebold2010}. Furthermore, when commutativity is assumed,  the mKdV and amKdV equations coincide, hence the B\"acklund chart as Fig. \ref{fig nc KdVchart} reduces to the Abelian B\"acklund chart in Fig. \ref{fig KdV-ext-chart} without the Dym equation. 
 
 On the other hand, also in the non-Abelian case, the invariance under the M\"obius group of transformations, suitably defined, allows to {\it double} the B\"acklund chart. 
 Hence, in both commutative as well as non-commutative cases new and/or well known invariances \cite{JMP2018} exhibited by the nonlinear evolution equations which appear the B\"acklund chart can be obtained, or, if already known, recovered. 

Furthermore, it should be noted that the Dym hierarchy is inserted in the Abelian B\"acklund chart.  Conversely,  at the moment, a non-commutative counterpart the reciprocal transformation remains to be investigated.  Nevertheless, solutions to non-commutative problems can be obtained \cite{Schiebold-6dic1, Schiebold-6dic2, Schiebold-6dic3, solutions2018}.

\subsection*{Acknowledgements}

The financial support of G.N.F.M.-I.N.d.A.M.,  I.N.F.N. and \textsc{Sapienza}  University of Rome, Italy are gratefully acknowledged.

\end{document}